\documentclass[12pt,a4paper]{JHEP}
\usepackage{amsmath,amsfonts,graphicx,vmargin,cite}

\title{Coherent Parton Showers\\ with Local Recoils}

\author{Simon Pl\"atzer, Stefan Gieseke \\
  Institut f\"ur Theoretische Physik\\ Universit\"at Karlsruhe, 
  76128 Karlsruhe, Germany}

\newcommand{\program}[1]{\textsf{#1}}

\preprint{HERWIG-09-06\\KA-TP-15-2009\\MCnet-09-15}
\abstract{ We outline a new formalism for dipole-type parton showers
  which maintain exact energy-momentum conservation at each step of the
  evolution.  Particular emphasis is put on the coherence properties,
  the level at which recoil effects do enter and the role of transverse
  momentum generation from initial state radiation.  The formulated
  algorithm is shown to correctly incorporate coherence for soft gluon
  radiation.  Furthermore, it is well suited for easing matching to
  next--to--leading order calculations.}

\begin{document}

\section{Introduction}

Parton shower simulation programs like \program{Pythia} and
\program{HERWIG} \cite{Corcella:2002jc,Sjostrand:2006za} have been the
workhorses for high energy physics experiments for a long time.  With
the advent of the Large Hadron Collider (LHC) at CERN many new
developments have been made in order to refine the existing programs and
to extend their applicability.  Currently, there are a few major multi
purpose event generators in use.  The \program{FORTRAN} generators have
been completely rewritten and extended as \program{Pythia8} and
\program{Herwig++} \cite{Gieseke:2003hm,Sjostrand:2007gs,Bahr:2008pv}
and the program \program{Sherpa} has been established
\cite{Gleisberg:2003xi,Gleisberg:2008ta}.  All these new generators come
with a great effort of new developments.

The classical parton showers with $1\to 2$ like branchings themselves
have been rewritten and reformulated in order to take into account mass
effects and improve the phase space coverage \cite{Gieseke:2003rz} or to
incorporate intertwined interactions with partons from additional hard
interactions in the underlying event \cite{Sjostrand:2004ef}.  Other
developments try to incorporate quantum interference effects or
subleading terms in the expansion in the number of colours
\cite{Nagy:2005aa,Nagy:2007ty,Nagy:2008ns,Nagy:2008eq}. Besides
many other efforts to improve the event simulation, e.g.\ in the area of
the underlying event
\cite{Sjostrand:2004pf,Skands:2009zm,Hoche:2007hg,Bahr:2008dy,
  Bahr:2008wk,Bahr:2009ek}, one of the most important questions that has
been addressed was for the matching with higher order matrix elements or
the merging of parton showers with multiple hard emissions at the tree
level.

The latter was formulated conceptually by Catani, Krauss, Kuhn and
Webber (CKKW) \cite{Catani:2001cc,Krauss:2002up} and, in a variant, by
L\"onnblad \cite{Lonnblad:2001iq} and implemented in various parton
shower programs
\cite{Schalicke:2005nv,Krauss:2005re,Lavesson:2005xu,Mrenna:2003if,diplsimon}.
An alternative approach to matching has been advocated by Mangano
\cite{Mangano:2002ea,Alwall:2007fs} and is known as MLM matching.
\cite{Alwall:2007fs} gives a comprehensive overview of current
implementations and shows detailed comparisons.  Recent improvements
address previous deficiencies of the CKKW algorithm in filling the phase
space \cite{Hoeche:2009rj,Hamilton:2009ne} that had been discussed
extensively in \cite{Lavesson:2007uu}.  Some problems have been overcome
with the help of so--called truncated showers that were introduced
earlier in the context of matching with next--to--leading order (NLO)
matrix elements \cite{Nason:2004rx}.  Merging with NLO matrix elements
has been recently studied in \cite{Lavesson:2008ah}.

First attempts to match parton showers and matrix elements at NLO have
been made in a phase space slicing approach
\cite{Dobbs:2001gb,Potter:2001ej,Potter:2000an} that suffered from some
(numerically small) systematic inconsistencies.  The more general and
systematic approach has been MC@NLO
\cite{Frixione:2002ik,Frixione:2003ei} that has since been extended to
include many processes and all possible colour structures
\cite{Frixione:2005vw,Frixione:2007zp,Frixione:2008yi}.  As the MC@NLO
is closely tied to a particular parton shower algorithm, the subtraction
terms that guarantee the consistent matching with the NLO matrix element
contributions have to be calculated once for a specific parton shower
program.  After all initial efforts have been made to match with
\program{HERWIG}, later also several processes have been matched to
\program{Herwig++}
\cite{LatundeDada:2007jg,LatundeDada:2009rr,Papaefstathiou:2009sr}.

While the MC@NLO approach is very successful and was developed for many
processes it may be considered to be tied too strongly to the underlying
parton shower and to suffer from negative weighted events, which, in
practice, never pose a real problem.  An alternative approach, now known
as \program{POWHEG}
\cite{Nason:2004rx,Nason:2006hfa,Frixione:2007nu,Frixione:2007vw}, has
been formulated by Nason \cite{Nason:2004rx}.  Here, the matching
formalism is based on a modified Sudakov form factor that contains the
real emission matrix element. This guarantees that the first emission of
the parton shower is the hardest one as well and can, therefore, be
described by the full matrix element.  In principle, this approach is
closely related to the so called hard matrix element corrections
\cite{Seymour:1994df,Platzer:2009b}.  This approach has been further
developed into a systematic matching scheme and applied to several
processes \cite{Alioli:2008gx,Alioli:2008tz} and was also widely used by
other groups, e.g.\ the \program{Herwig++} collaboration
\cite{LatundeDada:2006gx,LatundeDada:2008bv,Papaefstathiou:2009sr,Hamilton:2008pd,Hamilton:2009za}.

The subtraction based original MC@NLO approach is following the
requirement that the underlying parton shower algorithm must not be
modified.  Already shortly after the publication of the MC@NLO approach
it was, however, noted that the intrinsic subtraction in this scheme
could be simplified tremendously if the parton shower would follow
closely the subtraction terms that are used to regularize the soft and
collinear divergences in the NLO calculation to be matched.  This was
highlighted in \cite{Nagy:2005aa}, where also more formal developments
towards new partons were carried out.  Therefore, some groups started to
write new parton showers based on subtraction terms.  The groups
\cite{Schumann:2007mg,Dinsdale:2007mf} used the Catani--Seymour (CS)
\cite{Catani:1996vz} terms, while \cite{Winter:2007ye} based a shower on
Lund dipoles and \cite{Giele:2007di} on antenna subtraction terms.

In this paper we take the same viewpoint and present new theoretical
development towards a parton shower, based on CS subtraction kernels as
well.  After the question for soft coherence effects has been raised by
\cite{Dokshitzer:2008ia} in the context of a simplified toy model, we
would like to address this question for a full CS like parton shower.
The question of collinear radiation and the implication of the DGLAP
evolution was discussed in \cite{Nagy:2009re,Skands:2009tb}, not the
coherence properties, though.  In this paper, we formulate a parton
shower, based on CS subtraction kernels, in a detail that specifies a
full implementation.  We show that, with the right choice of evolution
variable and initial conditions it is indeed possible to find the
correct soft anomalous dimensions with a parton shower, based on CS
dipoles.  Therefore, it is possible to implement such a shower while
incorporating soft colour coherence effect in a way that has always been
a vital ingredient of the \program{HERWIG} and \program{Herwig++}
programs.
 
\section{Local Recoils, Form Factors and Coherence}

We consider a single parton emission off a pair of
partons with momenta $p_{ij}$ and $p_k$.
The probability for this emission is taken to be the sum
of two splitting functions, each associated with one leg. Using DGLAP
splitting kernels and the Sudakov decomposition for the splitting $p_{ij}\to q_i,q_j$,
\begin{eqnarray}
q_i &=& z p_{ij} + \frac{p_\perp^2}{2p_{ij}\cdot n\ z}n+k_\perp \ , \\
q_j &=& (1-z) p_{ij} + \frac{p_\perp^2}{2p_{ij}\cdot n\ (1-z)}n-k_\perp \ ,
\end{eqnarray}
where $k_\perp^2 = -p_\perp^2$ and $k_\perp\cdot p_{ij} = k_\perp\cdot n
= 0$ constitutes the usual collinear approximation, which may be
extended to the quasi-collinear approximation for emissions off massive
partons, \cite{Gieseke:2003rz}.  The light-like vector $n$ defines the
collinear direction, and therefore is used as the gauge vector in a
light-cone gauge when deriving the collinear-singular behaviour of QCD
matrix elements.  $n$ needs to be chosen along the colour connected
partner $p_k$, the so-called physical gauge, in which interference
diagrams are collinearly subleading such that the unregularized
splitting kernels are given by cut self-energy diagrams only.\footnote{We note
that this is a gauge choice for each singular limit of interest. The
definition of 'colour-connected' here applies in the large-$N_c$ limit but
may be generalized by including the full colour correlations present
at finite $N_c$.}

Note that, within this parametrization, the DGLAP splitting kernels are
functions of
\begin{equation}
z = \frac{n\cdot q_i}{n\cdot p_{ij}} \ .
\end{equation}
Indeed, there is not a single choice of light-cone gauge, but rather
a class of gauge choices which are connected by rescaling the gauge vector
$n$ ({\it i.e.} longitudinal boosts along the collinear direction),
for which the splitting kernels are left invariant.

We are interested in extending this picture such as not to perform an
approximation in the choice of kinematics, thereby introducing exact energy-momentum
conservation within the splitting $p_{ij},p_k\to q_i,q_j,q_k$. The choice of the recoil
strategy is not unique.
However, choosing a spectator to absorb the longitudinal
recoil of the splitting,
\begin{equation}
n=p_{k}\qquad q_k = \left(1-\frac{p_\perp^2}{2 p_{ij}\cdot p_k\ z(1-z)}\right) p_k
\end{equation}
is the only choice compatible with the remaining gauge degrees of freedom
in the functional form of the splitting kernels. As we shall also see,
this is the only choice which guarantees that the splitting functions
in a physical gauge do reproduce the correct soft behaviour.

\subsection{DGLAP Kernels, `Soft Correctness' and Angular Ordering}

As we are primarily interested in soft gluon radiation, we neglect
gluon splittings into quark-antiquark pairs in this section.

For final state radiation the spin-averaged DGLAP kernels are given by
\begin{equation}
P_{qg}(z) = C_F\left(\frac{2z}{1-z} + (1-z)\right)\ ,\qquad 
P_{gg}(z) =  2 C_A\left( \frac{z}{1-z} + \frac{1-z}{z} + z(1-z) \right) \ ,
\end{equation}
such that matrix elements squared, summed over all collinear
configurations factorize as
\begin{multline}
{}_{n+1}\langle {\cal M}(q_1,...,q_{n+1}) | {\cal M}(q_1,...,q_{n+1}) \rangle_{n+1} \to\\
\sum_{i=1}^n\sum_{j\ne i}\frac{4\pi \alpha_s}{q_i\cdot q_j} P_{ij}(z) \ 
{}_n\langle {\cal M}(q_1,...,p_{ij},...,q_{n+1}) | {\cal M}(q_1,...,p_{ij},...,q_{n+1}) \rangle {}_n \ .
\end{multline}
Note that in writing this expression, we do need to include a symmetry factor of $1/2$
along with the gluon splitting function.

As each amplitude $|{\cal M}\rangle$ is a colour singlet, {\it i.e.}
\begin{equation}
\sum_{i=1}^n {\mathbf T}_i^2 + \sum_{i=1}^n\sum_{j\ne i} {\mathbf T}_i\cdot {\mathbf T}_j = 0
\end{equation}
we may rewrite collinear factorization within the choice of the physical gauge
for single collinear configurations as
\begin{multline}
{}_{n+1}\langle {\cal M}(q_1,...,q_{n+1}) | {\cal M}(q_1,...,q_{n+1}) \rangle_{n+1} \to\\
\sum_{i=1}^n\sum_{j,k\ne i}\frac{4\pi \alpha_s}{q_i\cdot q_j} P_{ij}(z)|_{n=p_k} \ 
{}_n\langle {\cal M}(q_1,...,p_{ij},...,q_n) | {\mathbf C}_{(ij)k} | {\cal M}(q_1,...,p_{ij},...,q_n) \rangle_n \ ,
\end{multline}
where
\begin{equation}
{\mathbf C}_{ij} = -\frac{{\mathbf T}_i\cdot {\mathbf T}_j}{{\mathbf T}_i^2}
\end{equation}
is the colour correlation operator as introduced in \cite{Catani:1996vz}.

Within this framework, we have that
\begin{equation}
\frac{1}{q_i\cdot q_j}\left. \frac{z}{1-z}\right|_{n=p_k} = \frac{q_i\cdot p_k}{q_i\cdot q_j\ q_j\cdot p_k} \qquad
\frac{1}{q_i\cdot q_j}\left. \frac{1-z}{z}\right|_{n=p_k} = \frac{q_j\cdot p_k}{q_j\cdot q_i\ q_i\cdot p_k}
\end{equation}
such that the {\it single} splitting function $P_{ij}(z)|_{n=p_k}$
constitutes the complete, correct soft behaviour for the dipole $i,k$.
Note that the eikonal parts -- as well as any other part of a splitting
function -- is invariant under rescaling of the spectator momentum
$p_k$, which is an even stronger motivation to use the longitudinal
recoil strategy defined above.

This will also be a necessary requirement when trying to remove what we
call 'soft double counting'. As we will show now, this is closely
related to the coherence properties and logarithmic accuracy of a
particular shower setup.  To be precise, we consider the form factor
$\Delta_{ik}(Q^2,\mu^2)$ associated to a final-final dipole $i,k$ when
evolving from a hard scale $Q^2$ to a soft scale $\mu^2$. Regarding the
leading- (double) and next-to-leading (single) logarithmic contributions,
$\alpha_s^{n}L^{2n}$ and $\alpha_s^{n}L^{2n-1}$ with $L=\ln(Q^2/\mu^2)$
the correct behaviour can be obtained from the coherent branching
formalism \cite{Marchesini:1983bm}, reproducing the results of soft gluon resummation,
\cite{Bassetto:1984ik}, by considering the leading behaviour of the $z$-integrated
splitting kernel for $\mu^2\ll p_\perp^2\ll Q^2$. The resulting form factor reads
\begin{equation}
  -\ln \Delta_{ik}(Q^2,\mu^2) = 
  \int_{\mu^2}^{Q^2} \frac{{\rm d}p_\perp^2}{p_\perp^2}\frac{\alpha_s(p_\perp^2)}{2\pi}
  \left(\Gamma_i(p_\perp^2,Q^2)+\Gamma_k(p_\perp^2,Q^2)\right)\ ,
\end{equation}
where the Sudakov anomalous dimensions $\Gamma_k(p_\perp^2,Q^2)$ are given by
\begin{eqnarray}
  \label{eq:gammaq}
  \Gamma_q(p_\perp^2,Q^2) &=& C_F\left(\ln\frac{Q^2}{p_\perp^2} - \frac{3}{2}\right) \ , \\
  \label{eq:gammag}
  \Gamma_g(p_\perp^2,Q^2) &=& C_A\left(\ln\frac{Q^2}{p_\perp^2} - \frac{11}{6}\right) \ ,
\end{eqnarray}
receiving contributions both at the LL level from soft collinear, at the
NLL level from hard collinear radiation. Note that the latter, {\it i.e.}
the non-logarithmic terms in $\Gamma$ are determined by the average of the
soft-suppressed, $z$-regular terms of the splitting functions.

\subsection{Recoils and Soft Coherence}

We now want to include the effects of a finite recoil.  Within the
minimal recoil strategy outlined above this only affects the phase space
measure,
\begin{equation}
\frac{{\rm d}p_\perp^2}{p_\perp^2}{\rm d}z \to
\frac{{\rm d}p_\perp^2}{p_\perp^2}{\rm d}z \left(1-\lambda \frac{p_\perp^2}{z(1-z)s_{ik}}\right)\ ,
\qquad s_{ik}=2p_i\cdot p_k
\end{equation}
where we introduced $\lambda\to 1$ to explicitly keep track of these
effects.  Choosing a phase space region related to an ordering in
virtuality or transverse momentum,
\begin{equation}
4\mu^2 < \frac{p_\perp^2}{z(1-z)} < Q^2\ ,
\end{equation}
we find
\begin{eqnarray}
\Gamma^V_q(p_\perp^2,Q^2) &=& C_F\left(2\ln\frac{Q^2}{p_\perp^2} - \frac{3}{2} - 2\lambda\frac{Q^2}{s_{ik}}\right) \ , \\
\Gamma^V_g(p_\perp^2,Q^2) &=& C_A\left(2\ln\frac{Q^2}{p_\perp^2} - \frac{11}{6} - 2\lambda\frac{Q^2}{s_{ik}} \right) \ .
\end{eqnarray}
Note that here, the recoil effects enter at the level of next-to-leading
logarithms and the coefficient of the leading logarithms turns out to be
twice the correct result. The latter observation has been noted since
long \cite{Marchesini:1983bm}.  From this example it is very clear that
the simple fact that the DGLAP splitting functions reproduce the correct
soft behaviour is not enough for the correct soft anomalous dimension.
The wrong coefficient of the leading logarithmic contributions may be
attributed to a double counting of soft emissions, originating from the
fact that the above chosen phase space region does introduce an overlap
of the phase space available for emissions off either parton of the
dipole.

Choosing angular ordering in the variable $\tilde{q}$ by disentangling
soft and collinear limits\footnote{Note that in the soft limit(s), $z\to \epsilon\ ,\ 1-\epsilon$,
$p_\perp^2$ scales as $\epsilon^2$.}, and imposing phase space constraints
through a cutoff on the transverse momentum in the soft limit(s),
\begin{equation}
\tilde{q}^2=\frac{p_\perp^2}{z^2(1-z)^2} \qquad \mu^2 < z^2\tilde{q}^2\ ,\ (1-z)^2\tilde{q}^2
\qquad \tilde{q}^2 < Q^2
\end{equation}
we recover the correct anomalous dimensions (\ref{eq:gammaq},
\ref{eq:gammag}) with recoil effects entering beyond NLL,
\begin{eqnarray}
\Gamma^{AO}_q(p_\perp^2,Q^2) &=& C_F\left(\ln\frac{Q^2}{p_\perp^2} - \frac{3}{2}\right)
+C_F\frac{p_\perp}{Q}\left(1-2\lambda \frac{Q^2}{s_{ik}}\right) + {\cal O}\left(\frac{p_\perp^2}{Q^2}\right) \ , \\
\Gamma^{AO}_g(p_\perp^2,Q^2) &=& C_A\left(\ln\frac{Q^2}{p_\perp^2} - \frac{11}{6}\right) 
+2 C_A\frac{p_\perp}{Q}\left(1-\lambda \frac{Q^2}{s_{ik}}\right) + {\cal O}\left(\frac{p_\perp^2}{Q^2}\right) \ ,
\end{eqnarray}
the subleading terms giving rise to power corrections in the form
factor exponent.

Apart from the recoil effects, this result has a straightforward
explanation: The phase space region chosen for the angular ordered
evolution provides approximately disjoint regions (exactly disjoint in
the case of \program{Herwig++}) for emissions off either leg of the
dipole, thereby removing the soft double counting observed earlier.
Note that this observation would then in principle allow to include
local recoils within the angular ordered DGLAP evolution.

\subsection{Catani-Seymour Kernels and New Formalism}

As outlined in the previous sections, taking a minimal choice to treat
recoils yields a dipole-type picture. Within such a cascade it is
however difficult to maintain the strong angular ordering, which is tied
to the $1\to 2$ nature of independent jet evolution.

Choosing the phase space to be restricted by a cutoff on the transverse
momentum, thereby assuming an ordering in $p_\perp$ (or virtuality) is a
much more natural picture to consider for a dipole-type evolution. In
addition, this also removes complications when implementing matrix
element corrections, either stand alone or for the purpose of
\program{POWHEG}-type NLO matching as the first emission off a dipole
then is indeed the hardest emission.

To cure the problem of soft double counting generated by this evolution,
one may modify the splitting functions and 'continue' them over the whole
available phase space in such a way, that the soft-singular pieces reproduce
the correct soft behaviour when adding both modified splitting functions.

More precisely, for each leg $i$ we replace the eikonal part
by the radiation pattern associated with collinear emissions of $p_i$
\begin{equation}
\frac{p_i\cdot p_j}{p_i\cdot q\ p_j\cdot q} \to \frac{p_i\cdot p_j}{p_i\cdot q\ (p_i+p_j)\cdot q}
\end{equation}
while keeping the collinear parts exactly.  Note that this minimal
construction, which does not modify the singular properties following
from QCD, is nothing but the construction prescription for the
subtraction kernels introduced in \cite{Catani:1996vz}.  This picture of
local recoils using a single spectator parton is ideally supplemented
with exact factorization of the phase space considering no kinematic
approximation. One choice, which so far has been implemented
\cite{Schumann:2007mg,Dinsdale:2007mf} is to invert the kinematic
mappings as derived in \cite{Catani:1996vz}.

For initial state radiation, taking the Catani-Seymour factorization
literally does have shortcomings.  Most prominently, the choice of
keeping the initial state emitter's momentum collinear to the one before
emission leads for example to the fact that a final state singlet as in
Drell-Yan lepton pair production, does receive a non-vanishing
transverse momentum from the very first shower emission only. Further,
an initial-initial system emitting a parton left the spectator parton
unchanged, which might not be sufficient for the description of the
transverse momentum spectrum of the whole final state. The aim of this
work is to provide a formalism, which does overcome these
problems. Further, we are interested in the logarithmic accuracy 
and ordering of soft gluon radiation in our
setup reflecting coherence properties.

Starting from the final-state parametrization given above, the outline
of our formalism is as follows: We obtain a parametrization of the
kinematics for initial state emitters and/or spectators by considering
the physical splitting processes while maintaining exact energy-momentum
conservation {\it locally} to each branching, i.e. involving the
emitter-emission system and a single spectator only. The spectator is
restricted to take the longitudinal recoil of the splitting only. For initial state
radiation we do allow each initial state emission to generate transverse
momentum of the emitting incoming parton in a backward evolution. This
transverse momentum is then migrated to the complete final state system
by realigning the incoming partons to the beam axes at the end of the
evolution.

For final-final dipoles, we find that the anomalous dimensions take the
correct form apart from the fact, that the dependence on the arbitrary hard scale
$Q^2$ is being replaced by the dipole's invariant mass $s_{ik}$,
\begin{eqnarray}
\Gamma^{CS}_q(p_\perp^2,\cdot) &=& C_F\left(\ln\frac{s_{ik}}{p_\perp^2} - \frac{3}{2}\right)
-C_F\pi \lambda \frac{p_\perp}{\sqrt{s_{ik}}} + {\cal O}\left(\frac{p_\perp^2}{Q^2}\right) \ , \\
\Gamma^{CS}_g(p_\perp^2,\cdot) &=& C_A\left(\ln\frac{s_{ik}}{p_\perp^2} - \frac{11}{6}\right) 
-C_A\pi \lambda \frac{p_\perp}{\sqrt{s_{ik}}} + {\cal O}\left(\frac{p_\perp^2}{Q^2}\right) \ ,
\end{eqnarray}
with recoil effects entering beyond NLL.

We note that, in case of DGLAP kernels, the correct coefficient
of the leading logarithmic contributions
to the anomalous dimension is governed by the choice of boundaries
on the momentum fraction for a given (but arbitrary) hard scale $Q^2$,
\begin{equation}
\int_{0}^{1-\sqrt{\kappa}} \frac{{\rm d}z}{1-z} =\frac{1}{2} \ln\left(\frac{Q^2}{p_\perp^2}\right) \ ,
\end{equation}
with $\kappa=p_\perp^2/Q^2$.

The above findings for the anomalous dimension can essentially be traced back to
the fact that the transition from a DGLAP kernel possessing a soft singularity
$\sim 1/(1-z)$ to the appropriate Catani-Seymour kernel (while keeping track
of all recoil effects, i.e. considering the soft limit at fixed $p_\perp^2$)
is the simple replacement
\begin{equation}
\frac{1}{1-z} \to \left(1-\frac{\kappa_{ik}}{(1-z)}\right)\frac{1-z}{(1-z)^2+\kappa_{ik}} \ ,
\end{equation}
where $\kappa_{ik}=p_\perp^2/s_{ik}$. Here, the first factor is the effect of the
finite recoil stemming from the exact factorization of the phase space measure.

Within the variables to be outlined in detail in the next section, we find that this
pattern generalizes to the cases of initial state emitter or spectator partons, up
to a sign on the recoil term owing to timelike or spacelike virtualities of the
emitter or whether the relevant dipole scale is a spacelike momentum transfer
or invariant mass.

Choosing the $z$ boundaries (in the approximation considered above) to be given by 
\begin{equation}
z < 1-\frac{p_\perp^2}{Q^2} = 1-\kappa
\end{equation}
it is evident that the recoil contribution only gives rise to power corrections, while
the logarithmic contribution is given by
\begin{equation}
\frac{1}{2}\int_{\kappa^2}^1\frac{{\rm d}\xi}{\xi+\kappa_{ik}} = \frac{1}{2}\ln\left(\frac{s_{ik}}{p_\perp^2}\right)
+ \text{power corrections} \ ,
\end{equation}
thereby reproducing the correct coefficient up to the disappearance of the
arbitrary hard scale, an immediate consequence of the screening of the soft singularity
at fixed transverse momentum.

\subsection{Structure of the Evolution}

For final state radiation with final state spectator, our findings of
the previous section immediately signal a choice of the hard scale for a
single dipole originating from a hard process.  Choosing an arbitrary
hard scale $Q^2\ne s_{ik}$ will immediately result in the appearance of
spurious logarithmic contributions when performing the $p_\perp^2$
integration.

For example, at fixed $\alpha_s$ the leading logarithmic contributions for
a dipole $i,k$, with Casimir operators $C_{i,k}$ associated to the partons,
take the form
\begin{equation}
-\ln \Delta_{ik} = \frac{\alpha_s}{4\pi} (C_i^2+C_k^2)\ln\left(\frac{Q^2}{\mu^2}\right)
\ln\left(\frac{s_{ik}^2}{\mu^2Q^2}\right) + \text{NLL}
\end{equation}
instead of the expected result
\begin{equation}
-\ln \Delta_{ik} = \frac{\alpha_s}{4\pi} (C_i^2+C_k^2)\ln^2\left(\frac{Q^2}{\mu^2}\right) 
 + \text{NLL}\ ,
\end{equation}
the mismatch being manifest as an ambiguity at the level of next-to-leading logarithms.
We are therefore lead to the choice $Q^2=s_{ik}$, {\it i.e.} the hard scale associated
to a dipole is the respective invariant mass.

For initial state emitter or spectator partons, we assume that this generalizes
to choosing the hard scale in such a way as to fill the complete phase space,
modulo the infrared cutoff.

Note that this choice does not determine the ordering {\it per se}, but
only the choice of hard scale and the shape of the phase space
restriction when evolving between two scales. The ordering is to be
chosen in such a way, that the leading effects of multiple emissions off
each leg of the dipole do exponentiate. Due to the structure of the
splitting kernels given above and the additional complications from all
finite recoil effects the explicit exponentiation is beyond the scope of
this paper. 

Having however observed that we can reproduce the correct Sudakov
anomalous dimension, while avoiding soft double counting we additionally
note that within the variables chosen
\begin{equation}
p_\perp^2 = 2\frac{p_i\cdot q\ q\cdot p_k}{p_i\cdot p_k}
\end{equation}
for emission of a gluon of momentum $q$ off a dipole $(i,k)$. Ordering
emissions in this variable therefore corresponds to an ordering
reproducing the most probable history of multiple gluon emission
according to the eikonal approximation in the limit of soft gluons
strongly ordered in energy.

We therefore conclude that branchings within the physical kinematics
outlined above and based on the corresponding CS dipole splitting
functions allow us to construct a parton shower that has the right
coherence properties.  The final state emissions should in this case be
taken as outlined above, i.e.\ with the hard scale of a single cascade
chosen to be the dipole invariant mass and the evolution should be
strictly ordered in transverse momentum.  However, in the naive adoption
of the CS picture to a parton shower not every initial state emission
would contribute to the final state transverse momentum.  We will
formulate a more suitable approach below.

\section{Kinematics, Phase space and Splitting Probabilities}

The purpose of this section is to provide our new formalism in all
details, particularly the kinematic parametrization, phase space
convolution properties and boundaries, and the splitting probabilities
entering the evolution.  The explicit expressions for the splitting
kernels can be inferred from \cite{Catani:1996vz}, as we express the
variables chosen there in terms of the physical variables chosen for the
shower evolution.

\subsection{Final State Radiation}

\subsubsection{Final State Spectator}

Final state radiation with a final state spectator does represent
the generic version of the splitting kinematics chosen here. For a
splitting $(p_i,p_j)\to (q_i,q,q_j)$ we choose the standard Sudakov
decomposition
\begin{eqnarray}
q_i & = & z p_i +\frac{p_\perp^2}{z s_{ij}}p_j + k_\perp \\
q & = & (1-z) p_i +\frac{p_\perp^2}{(1-z) s_{ij}}p_j - k_\perp \\
q_j & = & \left(1-\frac{p_\perp^2}{z(1-z)s_{ij}}\right)p_j \ ,
\end{eqnarray}
where $k_\perp^2=-2p_i\cdot p_j \equiv -s_{ij}$ and $k_\perp\cdot p_{i,j}=0$.
The transverse momentum is defined in the dipole's rest frame to be purely spacelike,
\begin{equation}
\hat{p}_{i,j} = \left(\frac{\sqrt{s_{ij}}}{2},\pm\mathbf{p}\right)\ , \qquad
\hat{k}_\perp = \left(0,\mathbf{p}_\perp\right)\ , \qquad \mathbf{p}\cdot \mathbf{p}_\perp = 0 \ .
\end{equation}
Note that this does preserve the momentum of the emitting system,
$q_i+q+q_j=p_i+p_j$.
The parametrization gives rise to the phase space factorization \cite{Catani:1996vz}
\begin{equation}
{\rm d}\phi_3(q_i,q,q_j|Q)={\rm d}\phi_3(p_i,p_j|Q) 
\frac{1}{16\pi^2}\frac{{\rm d}\phi}{2\pi} {\rm d}p_\perp^2 
\frac{{\rm d}z}{z(1-z)}\left(1-\frac{p_\perp^2}{z(1-z)s_{ij}}\right)
\end{equation}
Note that, in the collinear limit, this is the
massless version of the kinematics as chosen in \cite{Gieseke:2003rz}.
It further constitutes the inversion of the 'tilde'-mapping, where the
variables $y$ and $z$ chosen in \cite{Catani:1996vz} are given by
\begin{equation}
y=\frac{p_\perp^2}{z(1-z)s_{ij}} \ ,\qquad z = \frac{p_j\cdot q_i}{p_i\cdot p_j} \ .
\end{equation}

The allowed phase space region is obtained by considering the limits on the
emitter's virtuality,
\begin{equation}
4\mu^2 < \frac{p_\perp^2}{z(1-z)} < Q_{\text{max}}^2 = s_{ij}
\end{equation}
such that
\begin{equation}
\mu^2 < p_\perp^2 < \frac{Q_{\text{max}}^2}{4}\ ,\qquad 
z_\pm = z_{\pm}=\frac{1}{2}\left( 1\pm \sqrt{1-\frac{4 p_\perp^2}{Q_{\text{max}}^2}} \right) \ .
\end{equation}

\begin{figure}
\centering
\includegraphics{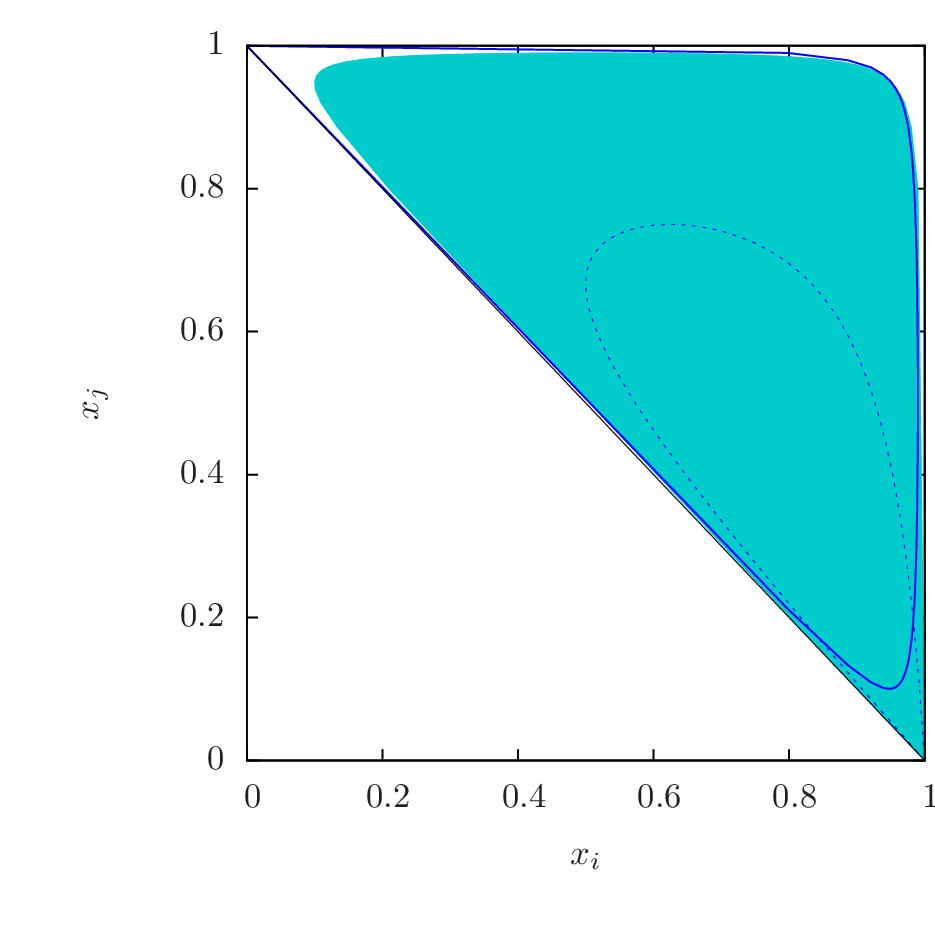}
\caption{Allowed phase space regions for emissions from a final-final
  dipole expressed in the Dalitz variables $x_k=2Q\cdot p_k/Q^2$ for a
  dipole of mass $s_{ij}=100\ {\rm GeV}$ and infrared cutoff $\mu=5\
  {\rm GeV}$. The shaded region is accessible for emissions off the
  parton $i$, whereas the area enclosed by the solid line is accessible
  for emissions off parton $j$. The area enclosed by the dotted line
  is an example of the phase space excluded when starting at a scale lower
  than $s_{ij}$.
  Note that the infrared cutoff is
  exaggerated for illustrative purposes only. In practice, almost the
  whole physical phase space will be available.}
\end{figure}

Averaging over azimuth, the final-final splitting kernels take the form
\begin{equation}
\frac{8\pi \alpha_s}{2 q_i\cdot q}\langle V(p_\perp^2,z)\rangle
\end{equation}
such that the splitting probability is
\begin{equation}
{\rm d}P = \frac{\alpha_s}{2\pi}\langle V(p_\perp^2,z)\rangle
\left(1-\frac{p_\perp^2}{z(1-z)s_{ij}}\right)
\frac{{\rm d}p_\perp^2}{p_\perp^2} {\rm d}z \ .
\end{equation}
Note that, comparing to the collinear limit, the effect of finite recoils is to
act as a damping factor for large-angle hard emissions,
provided that $y<1$ which is a consequence of the phase space
boundary.

\subsubsection{Initial State Spectator}

For an initial state spectator we consider
the crossing $q_j\to -q_a$, $p_j\to -p_a$, such that
\begin{eqnarray}
q_i & = & z p_i +\frac{p_\perp^2}{z s_{ia}}p_a + k_\perp \\
q & = & (1-z) p_i +\frac{p_\perp^2}{(1-z) s_{ia}}p_a - k_\perp \\
q_a & = & \left(1+\frac{p_\perp^2}{z(1-z)s_{ia}}\right)p_a \ .
\end{eqnarray}
Note that exact momentum conservation is trivially implemented
by just the fact that the parametrization for a final state spectator
does respect this constraint.
The transverse momentum is defined be purely spacelike in a frame where,
\begin{equation}
\hat{p}_{i,a} = \left(\frac{\sqrt{s_{ia}}}{2},\pm\mathbf{p}\right)\ , \qquad
\hat{k}_\perp = \left(0,\mathbf{p}_\perp\right)\ , \qquad \mathbf{p}\cdot \mathbf{p}_\perp = 0 \ .
\end{equation}

The phase space measure then obeys the convolution
\begin{equation}
{\rm d}\phi^F_2(q_i,q|Q;P_a,q_a,x_a) =
{\rm d}\phi^F_1(p_i|Q;P_a,p_a,x x_a)\frac{{\rm d}\phi}{2\pi}
\frac{x}{16\pi^2}\frac{{\rm d}z}{z(1-z)}{\rm d}p_\perp^2 \ ,
\end{equation}
where
\begin{equation}
x=\frac{1}{1+\frac{p_\perp^2}{z(1-z)s_{ia}}}\ , \qquad z = \frac{p_a\cdot q_i}{p_a\cdot p_i}
\end{equation}
and it is straightforward to verify that this indeed gives rise to the
phase space convolution as given in \cite{Catani:1996vz}.
Including the parton distributions and the kinematic factor of the 
partonic flux, the relevant measure is
\begin{multline}
\frac{f_a(x_a)}{4 q_a\cdot n}{\rm d}\phi^F_2(q_i,q|Q;P_a,q_a,x_a) {\rm d}x_a
=\\ \left(\frac{f_a(x_a/x)}{f_a(x_a)}\theta(x-x_a)\frac{{\rm d}\phi}{2\pi}
\frac{x}{16\pi^2}\frac{{\rm d}z}{z(1-z)}{\rm d}p_\perp^2
\right) \frac{f_a(x_a)}{4 p_a\cdot n}{\rm
  d}\phi^F_1(p_i|Q;P_a,p_a,x_a) {\rm d}x_a \ .
\end{multline}
The phase space limits can be obtained as for the final state case,
\begin{equation}
4\mu^2 < \frac{p_\perp^2}{z(1-z)} < Q_{\text{max}}^2 \ ,
\end{equation}
where, owing to $x>x_a$, the hard scale of a dipole is now given by
\begin{equation}
Q_{\text{max}}^2 = s_{ia}\frac{1-x_a}{x_a} \ .
\end{equation}
Averaging over azimuth, the final-initial splitting kernels take the form
\begin{equation}
\frac{8\pi \alpha_s}{2 q_i\cdot q}\frac{1}{x}\langle V(p_\perp^2,z)\rangle
\end{equation}
such that the splitting probability is
\begin{equation}
{\rm d}P = \frac{\alpha_s}{2\pi}\langle V(p_\perp^2,z)\rangle
\frac{f_a(x_a/x)}{f_a(x_a)}\theta(x-x_a)
\frac{{\rm d}p_\perp^2}{p_\perp^2} {\rm d}z \ .
\end{equation}
Note that the finite recoil enters only in the PDF ratio, reproducing
the correct collinear limit when $x\to 1$. Once again, the effect of the
finite recoil is a damping of hard emissions for $x\sim x_a$.

\subsection{Initial State Radiation}

A construction of initial state radiation by just crossing prescriptions is
not obvious owing to the fact that the shower evolution is formulated
as a {\it backward} evolution.

The physical variables thus need to be defined from the physical {\it forward}
kinematics. For the physical emission process $q_a\to p_a,q$ the relevant Sudakov
decomposition for the emission momentum $q$ is
\begin{equation}
q_{\rm forward} = (1-z) q_a + \frac{p_\perp^2}{2 n\cdot q_a (1-z)} n - k_\perp \ ,
\end{equation}
where $n$ is the backward lightcone direction defining the collinear direction, {\it i.e.}
the final or initial state spectator's momentum.

The parametrization above is most conveniently inverted to backward evolution $p_a\to q_a,q$
by considering the process in a frame where $q_a=p_a/x$, giving rise to
\begin{equation}
q_{\rm backward} = \frac{(1-z)}{x} p_a + \frac{p_\perp^2}{2 n\cdot p_a (1-z)} n - \frac{1}{\sqrt{x}}k_\perp \ .
\end{equation}
We therefore define the Lorentz invariant physical variables to be given by
\begin{equation}
x\ q_a\cdot q = \frac{p_\perp^2}{1-z} \qquad x\ n\cdot q = (1-z) n\cdot p_a \ .
\end{equation}

The parametrization keeping the emitter aligned with the beam axis can then
be related to a parametrization where the initial state parton after (backward evolution)
emission does acquire a finite transverse momentum while keeping the spectator
after emission aligned with the one before emission.

It is this type of splitting kinematics which allows any emission off an initial state
parton to contribute transverse momentum to a final state system after having
applied a proper realignment boost once the parton shower evolution has
terminated. Ideally, this final boost should {\it not} be related to the
parametrization chosen but being determined in a process dependent way such as
to leave the interesting kinematic quantities of the hard process invariant.

\subsubsection{Final State Spectator}

For initial state emissions with final state spectator, $p_a,p_j\to q_a,q,q_j$, using
the variables introduced in \cite{Catani:1996vz}, 
\begin{equation}
x = \frac{p_a\cdot p_j}{(p_a-p_j)\cdot q_a} \qquad u = \frac{q_j\cdot q_a}{(p_a-p_j)\cdot q_a} \ ,
\end{equation}
we use the parametrization
\begin{eqnarray}
q_a & = & \frac{1-u}{x-u} p_a + \frac{u}{x}\ \frac{1-x}{x-u} p_j + \frac{1}{u-x}k_\perp \\
q & = & \frac{1-x}{x-u} p_a + \frac{u}{x}\ \frac{1-u}{x-u} p_j + \frac{1}{u-x}k_\perp \\
q_j & = & \left(1-\frac{u}{x}\right) p_j \ ,
\end{eqnarray}
which does preserve the momentum transfer, $q+q_j-q_a=p_j-p_a$. The transverse
momentum obeys
\begin{equation}
k_\perp^2 = -u(1-u)\frac{1-x}{x} s_{aj} \qquad s_{aj} = 2 p_a\cdot p_j \ .
\end{equation}
Considering the collinear limit $u\to 0$, it is evident that the relevant momentum fraction
is $x$ and we are therefore lead to choose the physical variables to be given by
\begin{equation}
u = \frac{\kappa}{1-z}\ , \qquad x = \frac{z(1-z)-\kappa}{1-z-\kappa}\ , \qquad \kappa = \frac{p_\perp^2}{s_{aj}} \ .
\end{equation}

Indeed, the Lorentz transformation
\begin{multline}
R^\mu {}_\nu =\\ \delta^\mu_\nu +
 \frac{x}{(1-u)(x-u)}\ \frac{k_\perp^\mu k_{\perp\nu}}{p_a\cdot p_j} + 
 \frac{u(1-x)}{x-u}\ \frac{K^\mu K_\nu}{p_a\cdot p_j} +
\frac{x}{x-u}\ \frac{k_\perp^\mu K_\nu-K^\mu k_{\perp\nu}}{p_a\cdot p_j}
\end{multline}
with $K=p_a+p_j$ relates the above parametrization to one preserving the
direction of the incoming parton,
\begin{eqnarray}
Rq_a & = & \frac{1}{x} p_a \\
Rq & = & u p_j +(1-u)\frac{1-x}{x} p_a -k_\perp \\
Rq_j & = & (1-u) p_j +u\frac{1-x}{x} p_a +k_\perp \ .
\end{eqnarray}

In order to derive the phase space convolution properties associated with the
parametrization given above, we employ the formalism outlined in the appendix.
Substituting
\begin{equation}
u = \frac{y}{w+y(1-w)}\ , \qquad x = \frac{1}{w+y(1-w)}
\end{equation}
the parametrization above is mapped to
\begin{eqnarray}
(-q_a) & = & w (-p_a) + (1-w)\ y\ p_j - q_\perp \\
q & = & (1-w) (-p_a) + w\ y\ p_j + q_\perp \\
q_j & = & \left(1-y\right) p_j \ ,
\end{eqnarray}
with $q_\perp^2=-s_{aj} y w (1-w)$ such that the generalized phase space measure factors as
\begin{equation}
{\rm d}\phi_3(q_j,q,-q_a|Q) = \frac{s_{aj}}{16\pi^2}\frac{{\rm d}\phi}{2\pi} 
\frac{{\rm d}x}{x^3}\ {\rm d}u\
{\rm d}\phi_2(p_j,-p_a|Q) \ .
\end{equation}

Having identified $x$ to be the relevant momentum fraction we consider
the hadronic collision in a frame where\footnote{Note that there is no a priori
relation between incoming hadron and parton momenta in our formulation.}
\begin{equation}
P_a\cdot q_a = \frac{1}{x} P_a\cdot p_a\ , \qquad N\cdot q_a = \frac{1}{x} N\cdot p_a \ ,
\end{equation}
such that the phase space convolution
properties of both parametrizations become equivalent at {\it hadron} level,
\begin{equation}
{\rm d}\phi^F_2(q_i,q|Q;P_a,q_a,x_a) =
{\rm d}\phi^F_1(p_i|Q;P_a,p_a,x x_a)\frac{{\rm d}\phi}{2\pi}
\frac{1}{16\pi^2}\frac{{\rm d}z}{z(1-z)-\kappa}{\rm d}p_\perp^2 \ ,
\end{equation}
We stress that the crucial difference is related to the fact that, considering
the physical forward evolution, our parametrization does generate a finite
transverse momentum for the parton entering the hard process after additional
parton emission.

Averaging over azimuth, the initial-final splitting kernels take the form
\begin{equation}
\frac{8\pi \alpha_s}{2 q_a\cdot q}\frac{1}{x}\langle V(p_\perp^2,z)\rangle
\end{equation}
such that the splitting probability is
\begin{equation}
{\rm d}P = \frac{\alpha_s}{2\pi}\langle V(p_\perp^2,z)\rangle
\frac{f_a(x_a/x)}{f_a(x_a)}\theta(x-x_a)
\frac{{\rm d}p_\perp^2}{p_\perp^2} \frac{(1-z){\rm d}z}{z(1-z)-\kappa} \ .
\end{equation}
Note that in the collinear limit, $\kappa\to 0$ we have $x\to z$
such that the collinear behaviour is properly reproduced.\footnote{For readability
we have suppressed indexing a possible flavour change of the incoming parton.}

The phase space boundaries are given by the requirement that $x_a<x$,
\begin{equation}
\mu^2 < p_\perp^2 < \frac{(1-x_a)s_{aj}}{4}\ , \qquad z_{\pm}=\frac{1}{2}\left(1+x_a \pm (1-x_a)
\sqrt{1-\frac{4 p_\perp^2}{(1-x_a)s_{aj}}} \right) \ .
\end{equation}

\begin{figure}
\centering
\includegraphics{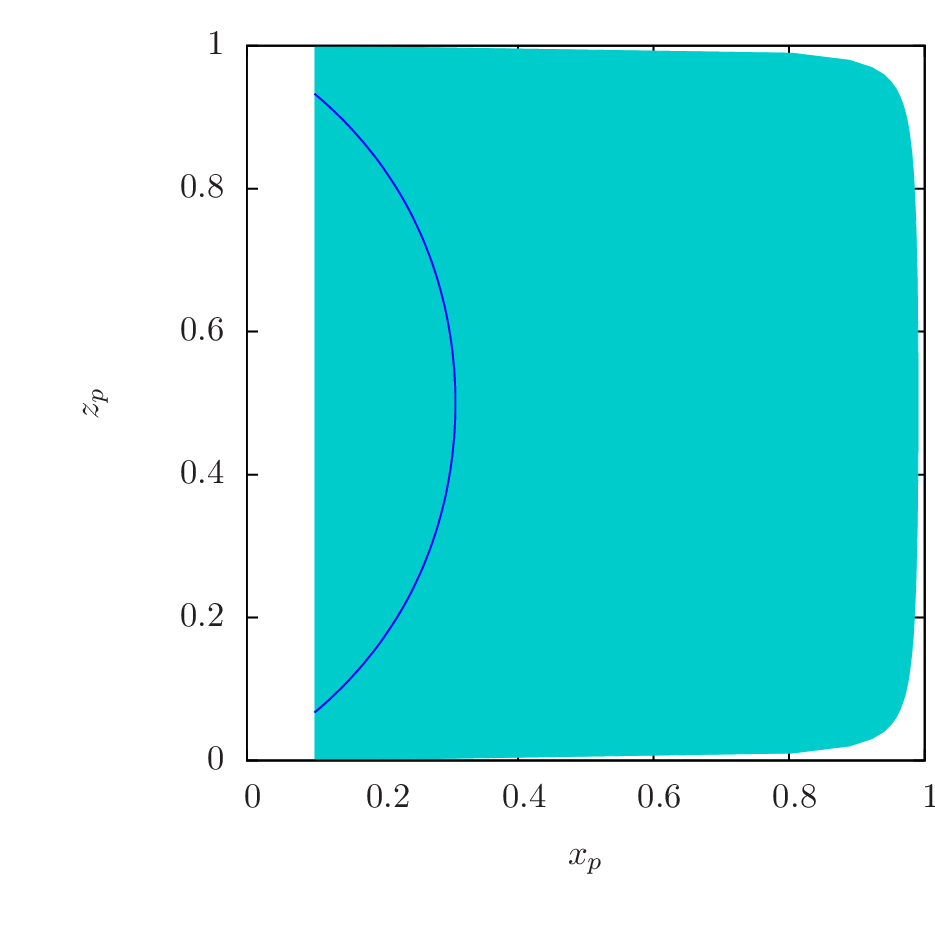}
\caption{Available phase space for a final-initial dipole with invariant momentum transfer
$\sqrt{s_{aj}}=\sqrt{-t}=100\ {\rm GeV}$ and an infrared cutoff of $5\ {\rm GeV}$. The shaded region
is accessible starting at the hard scale, the region enclosed by the solid line is an example 
of the phase space excluded when 
starting at a lower scale. The phase space regions for an initial-final dipole are identical.
For a final-initial dipole, the variables are $x_p=x$, $z_p=z$, for the initial-final one
$x_p=x$, $z_p=1-u$. Note that in the latter case $u\to 1$ and $u\to 0$ correspond to a collinear
limit.}
\end{figure}

\subsubsection{Initial State Spectator}

Initial state radiation with initial state spectator, $p_a,p_b\to q_a,q,q_b$ is described by
the parametrization
\begin{eqnarray}
q_a & = & \frac{1}{v+x} p_a + \frac{v}{x}\ \frac{1-v-x}{v+x} p_b + \frac{1}{v+x}k_\perp \\
q & = & \frac{1-v-x}{v+x} p_a + \frac{v}{x}\ \frac{1}{v+x} p_b + \frac{1}{v+x}k_\perp \\
q_b & = & \left(1+\frac{v}{x}\right) p_b \ ,
\end{eqnarray}
preserving $q-q_a-q_b=-p_a-p_b$.
The transverse momentum is defined to be purely spacelike in the dipole's rest frame
and obeys
\begin{equation}
k_\perp^2 = -(1-v-x)\frac{v}{x} s_{ab} \qquad s_{ab} = 2 p_a\cdot p_b \ .
\end{equation}
The variables $x$ and $v$ are those introduced in \cite{Catani:1996vz}, 
\begin{equation}
x = \frac{p_a\cdot p_b}{q_a\cdot q_b}\ , \quad v = \frac{q_a\cdot q}{q_a\cdot q_b} \ ,
\end{equation}
and we define the physical variables to be given by
\begin{equation}
x = \frac{z(1-z)-\kappa}{1-z}\ , \qquad v=\frac{\kappa}{1-z}\ , \qquad \kappa = \frac{p_\perp^2}{s_{ab}} \ .
\end{equation}
Note that the Lorentz transformation
\begin{multline}
S^\mu {}_\nu =\\ \delta^\mu_\nu +
\frac{p_b\cdot p_a}{p_b\cdot q_a\ q_a\cdot p_a}q_a^\mu q_{a,\nu} + 
\frac{p_b\cdot q_a}{p_b\cdot p_a\ q_a\cdot p_a}p_a^\mu p_{a,\nu}-
\frac{1}{q_a\cdot p_a}\left( q_a^\mu p_{a\nu}+ p_a^\mu q_{a\nu}\right)
\end{multline}
does transform this parametrization to a parametrization where 
\begin{equation}
Sq_a = \frac{1}{x+v}p_a\ , \qquad Sq_b = \frac{x+v}{x} p_b \ .
\end{equation}
Following the arguments of the previous section we then find the
phase space convolution
\begin{multline}
{\rm d}\phi^F_1(q|Q;P_a,q_a,x_a;P_b,q_b,x_b) =\\
{\rm d}\phi^F_0\left(\left.\right|Q;P_a,p_a,(x+v) x_a;P_b,p_b,\frac{x}{x+v} x_b\right)\frac{{\rm d}\phi}{2\pi}
\frac{1}{16\pi^2} \frac{{\rm d}z}{z(1-z)-\kappa}{\rm d}p_\perp^2\ .
\end{multline}
Averaging over azimuth, the initial-final splitting kernels take the form
\begin{equation}
\frac{8\pi \alpha_s}{2 q_a\cdot q}\frac{1}{x}\langle V(p_\perp^2,z)\rangle
\end{equation}
such that the splitting probability is
\begin{equation}
{\rm d}P = \frac{\alpha_s}{2\pi}\langle V(p_\perp^2,z)\rangle
{\cal F}_{ab}
\frac{{\rm d}p_\perp^2}{p_\perp^2} \frac{(1-z){\rm d}z}{z(1-z)-\kappa} \ ,
\end{equation}
with
\begin{equation}
{\cal F}_{ab} =
\frac{f_a(x_a/(x+v))}{f_a(x_a)}\theta(x+v-x_a) \frac{f_b(x_b(x+v)/x)}{f_b(x_b)}\theta\left(\frac{x}{x+v}-x_b\right)
\end{equation}
the ratio of incoming parton flux.
\footnote{As for the final state spectator, we have suppressed indexing a possible 
flavour change of the incoming parton.} Note that in the collinear limit,
$v,\kappa\to 0$ and $x\to z$ such that we find the correct collinear behaviour.

We remark that it would be possible to keep the spectator unchanged
upon properly substituting the integrations over the incoming momentum fractions
(the Jacobian being equal to one). This, however, would invalidate the fact
that the above given parametrization of the splitting kinematics does preserve
energy-momentum locally involving the emitter-emission-spectator system
only (in fact, after applying the relevant Lorentz transformation $S$
this would constitute the inversion of the kinematics as used in the
dipole subtraction context). A further argument to not keeping the spectator
unchanged is that, following the discussion on soft and collinear factorization,
we see no reason why the emission off the {\it colour connected} system $p_a,p_b$
should leave $p_b$ unchanged except for a strictly soft and/or collinear emission.

The phase space limits are now determined from $x>x_a x_b = \tau$ to be given by
\begin{equation}
z_\pm = \frac{1}{2}\left(1+\tau\pm(1-\tau)\sqrt{1-\frac{4 p_\perp^2}{(1-\tau)^2 s_{ab}}} \right) \ ,
\qquad p_\perp^2 < \frac{(1-\tau)^2 s_{ab}}{4} \ .
\end{equation}

\begin{figure}
\centering
\includegraphics{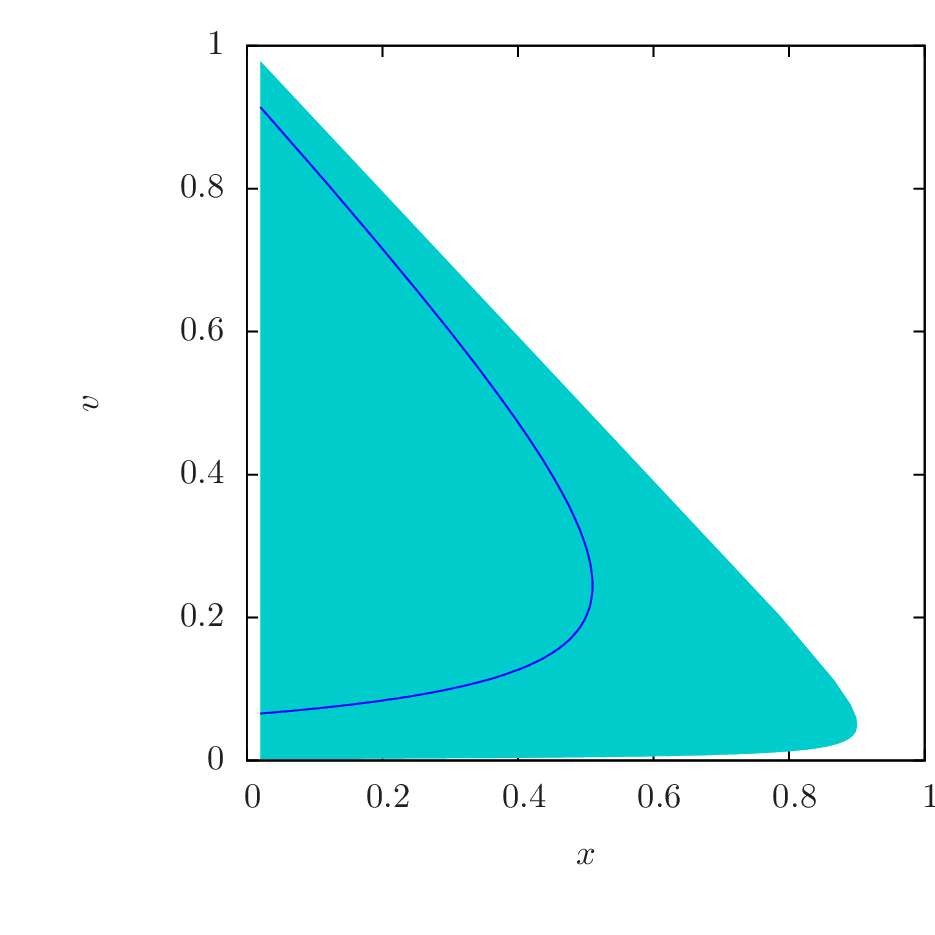}
\caption{Available phase space for emissions off an initial-initial dipole of mass $100\ {\rm GeV}$
with $\tau=0.02$ and infrared cutoff $5\ {\rm GeV}$. The shaded region is the available phase space when starting
from the hard scale, the region enclosed by the solid line is an example of the phase space excluded when
starting at a lower scale. }
\end{figure}

\section{Conclusions}

We have argued that parton showers, based on Catani--Seymour dipoles
have a number of properties that turn out to be useful when one
considers the matching with NLO matrix elements.  Several other groups
\cite{Schumann:2007mg,Dinsdale:2007mf} have already written a parton
shower program, following this motivation.  In these approaches,
however, the choices of phase space boundaries and evolution variable
were made rather intuitively.  The question for coherence properties,
in particular, was left open.

Taking this as a starting point we have investigated the soft
coherence properties of a CS parton shower formalism.  We are lead to
the transverse momentum of each dipole as the evolution variable of
our cascade.  Furthermore, we chose a hard scale that allows us to
access the whole kinematically allowed phase space.  We then
explicitly show that such a parton shower can reproduce the expected
Sudakov anomalous dimensions and hence include soft coherence effects
properly.

We have specified all details of such a parton shower that
will be important for its implementation.  In particular, we addressed
the issue of transverse momentum from initial state radiation that will
build up from several emissions in our case.  An implementation of such
a parton shower is underway.

\section*{Acknowledgements}

We are grateful to the other \program{Herwig++} authors and Leif
L\"onnblad for extensive collaboration.  We thank Y.\ Dokshitzer and M.\
Seymour for very fruitful discussions. 
This work was supported in part by the European Union Marie Curie
Research Training Network MCnet under contract MRTN-CT-2006-035606 and
the Helmholtz Alliance ``Physics at the Terascale''.

\appendix

\section{Treatment of Collinear Factorization}

In view of our probabilistic treatment of parton showers \cite{Platzer:2009b},
we choose to rephrase collinear factorization in a way that fits
to the picture of assigning the usual phase space measure
\begin{equation}
{\rm d}\phi_1(p)=\frac{{\rm d}^{3}{\mathbf p}}{(2\pi)^{3} 2 p^0}
\end{equation}
to {\it each} parton in the evolution, and not only final
state partons. In turn, this allows us to derive phase space
convolution properties using the general phase space factorization
inherent to the parametrization of final state splittings with
final state spectator.

We therefore rewrite
\begin{equation}
\int_0^1{\rm d}x_a = \int {\rm d}x_a \int \frac{{\rm d}^3{\mathbf p_a}}{(2\pi)^3 2 p_a^0}
\delta_F(P_a,p_a,x_a)
\end{equation}
where
\begin{equation}
\delta_F(P_a,p_a,x_a) = \frac{16\pi^2}{2 P_a\cdot P_b}\delta\left(\frac{P_a\cdot p_a}{P_a\cdot P_b}\right)
\delta\left(\frac{P_b\cdot p_a}{P_a\cdot P_b} - x_a\right)\theta(x_a(1-x_a)) \ .
\end{equation}
Here, $P_a$ denotes the lightlike momentum of the incoming hadron and the
lightlike vector $P_b$ defines the collinear direction, {\it i.e.} is taken to be
the momentum of the second incoming particle. We note that this extends
straightforwardly to dimensional regularization.

The phase space measure with an incoming parton then takes the form
\begin{equation}
{\rm d}\phi^F_{n}(p_1,...,p_n|Q;P_a,p_a,x_a) =
{\rm d}\phi_{n+1}(p_1,...,p_n,-p_a|Q)\delta_F(P_a,p_a,x_a)
\end{equation}
in terms of the general phase space measure
\begin{equation}
{\rm d}\phi_{n+1}(q_1,...,q_n|Q) = {\rm d}\phi_1(q_1)\cdots {\rm d}\phi_1(q_n) (2\pi)^4 \delta(q_1+\cdots+q_n-Q) \ .
\end{equation}

The phase space convolutions as given in \cite{Catani:1996vz} take the
form
\begin{equation}
\int_0^1{\rm d}x_a f(x_a) \int_0^1 {\rm d} z\ 
{\rm d}\phi_n(p_1,...,p_n | Q+z q_a)\ q_a\cdot p_i J(z)\delta(z-x){\rm d} x \ ,
\end{equation}
where $q_a=(1/z)p_a=x_a P_a$ is constrained by collinear factorization. It is then easy to prove that
the same convolution within our framework reads
\begin{equation}
\int {\rm d}x_a f(x_a)\int_0^1 \frac{{\rm d}z}{z} p_a\cdot p_i\ J(z)\delta(z-x) {\rm d}\phi_n^F(p_1,...,p_n|Q;P_a,p_a,z x_a)
{\rm d} x
\end{equation}
and gives rise to the factorization
\begin{equation}
\int{\rm d}x_a f(x_a)\left( \frac{f(x_a/x)}{f(x_a)}p_a\cdot p_i J(x)\theta(x-x_a) \frac{{\rm d} x}{x^2}\right)
{\rm d}\phi^F_n(p_1,...,p_n|Q;P_a,p_a,x_a) \ .
\end{equation}

\bibliography{dipoleshower}

\end{document}